# Isolation Support for Service-based Applications
# A Position Paper

Paul Greenfield
School of Information Technologies
University of Sydney NSW 2006
Australia
+61 2 9560 4952

p.greenfield@computer.org

Alan Fekete
School of Information Technologies
University of Sydney NSW 2006
Australia
+61 2 9351 4287

fekete@it.usyd.edu.au

Julian Jang
CSIRO ICT Centre
PO Box 76 Epping NSW 1710
Australia
+61 2 9372 4658

julian.jang@csiro.au

Dean Kuo
School of Computer Science
University of Manchester
UK M13 9PL
+44 (0) 161 275 0683

dkuo@cs.man.ac.uk

Surya Nepal
CSIRO ICT Centre
PO Box 76 Epping NSW 1710
Australia
+61 2 9372 4256

surya.nepal@csiro.au

## ABSTRACT
In this paper, we propose an approach that provides the benefits of isolation in service-oriented applications where it is not feasible to use the traditional locking mechanisms used to support ACID transactions. Our technique, called 'Promises', provides a uniform mechanism that clients can use to ensure that they can rely on the values of information resources remaining unchanged in the course of long-running operations. The Promises approach covers a wide range of implementation techniques on the service side, all allowing the client to first check a condition and then rely on that condition still holding when performing subsequent actions.

## Categories and Subject Descriptors
D.2.12 [**Software engineering**]: Interoperability; D.2.11 [Software Engineering]: Software Architectures.

## 1. INTRODUCTION
Web Services and service-oriented architectures are widely accepted as being the technologies that will be used to build the next generation of Internet-scale distributed applications. These applications are constructed by gluing together opaque and autonomous services, possibly supplied by business partners and third party service providers, to form loosely-coupled virtual applications. The services model is extremely simple but, unfortunately, this simplicity does not mean that service-based applications will prove to be easy to develop in practice, or be sufficiently reliable and robust.

Building robust large-scale stateful distributed systems is a long-standing and inherently hard problem. Some of the difficulties come as consequences of having to deal with the effects of concurrency and partial failures, and are made worse by the opaque and autonomous nature of services. Traditional distributed ACID transaction technologies provide an elegant and powerful solution to these problems, but depend on assumptions of trust and timeliness that no longer apply in the new loosely-coupled services-based world.

Our earlier work [4] on improving the robustness of service-based distributed applications focussed on the consistency problem: how to ensure that the set of autonomous services making up one of these applications always finish in consistent states despite failures, races and other such difficulties. Rather than attempting to provide the equivalent of traditional distributed transactions for the loosely-coupled Web Services world, our approach instead was to develop tools, programming models and protocols for the detection and avoidance of consistency faults, at both design time and at run-time. The key to this work was establishing a relationship between internal service states, messages and application-level protocols. This insight let us transform the problem of ensuring consistent outcomes into a protocol problem that could be addressed using proven techniques from the world of protocol verification. We then developed tools that could test whether the contracts defining the behaviour of two services were compatible and that their interactions would never lead to an inconsistent outcome. The same message-based definitions of correctness and consistency were also used as the basis for a protocol for dynamically checking for consistency failures at the termination of service-based applications, without requiring an overall coordinator or a global view of the entire application.

This earlier work addressed only the 'atomicity' part of the larger problem of simplifying the construction of robust and reliable service-based distributed applications. We could prove that the use of correctly designed contracts and the resulting application protocols could avoid inconsistent outcomes, but we still required the programmer to provide code to handle each possible message under every possible state. For example, the methodology of [4] requires a merchant service to have code for the situation where payment arrives for an accepted order when there is insufficient stock on hand. In the simpler world of ACID transactions, programmers could simply start a transaction and check stock levels when the order was accepted, and then rely on sufficient stock being available throughout the rest of the order process, regardless of any concurrent orders or other activities. The challenge we faced was providing a useful degree of isolation in a services-based world where autonomy and lack of trust meant that traditional lock-based isolation mechanisms could not be used. Our approach to this problem was to first identify a range of real-world examples where the lack of isolation was actually a problem, and then to understand and generalise the solutions to these problems already adopted in traditional business processes. The result of this work is a general pattern and protocol called 'Promises'.

## 2. PROMISES

A Promise is an agreement between a client application (a 'promise client') and a service (a 'promise maker'). By accepting a promise request, a service guarantees that some set of conditions ('predicates') will be maintained over a set of resources for a specified period of time.

In the conceptual model discussed in this paper, promises are granted and guaranteed by a Promise Manager rather than directly by services. A promise manager sits between clients and application services and implements Promise functionality on behalf of a number of services and resource managers. The job of a promise manager is to work with application services and resource managers to grant or deny promise requests, check on resource availability and ensure that promises are not violated.

Client applications can determine what resources they need to have available in order to always complete successfully, express these as a precise set of predicates and send them to the relevant promise manager as a promise request. The promise manager will examine both the complete set of existing promises and the availability of the requested resources, and either grant or reject the promise request. Once a promise request is granted, the client application is isolated from the effects of concurrent activities with respect to the resources protected by its promises. For example, the merchant order-handling process we mentioned above can now ask the manager of the stock resource for an initial promise that the goods required to meet an order will not be sold to anyone else for the duration of the order handling process. Once this promise has been obtained, the order-handling process can proceed with the knowledge that the required stock will be available when needed, even though concurrent order processes may be also selling the same type of goods to other customers.

Traditional lock-based isolation can be seen as a very strong and monolithic form of promise, one where the resource manager is guaranteeing that no other concurrent process can alter, or possibly even examine, the state of a protected resource for the duration of an operation. The proposed promise-based isolation mechanism is weaker but can be just as effective because it can be more precise. The predicates contained within a promise specify a client application's exact resource requirements, allowing other promises covering the same resources to be granted concurrently as long as they do not conflict with any already granted promises.

Promises do not last forever. The client and promise manager agree on the period of time for which a promise will be valid as part of the promise request/granting process, and promises will expire at the end of this time. Promise managers return 'promise-expired' errors to clients that attempt to perform operations under the protection of expired promises.

Promise-aware applications can be written with the knowledge that the resources they need for successful completion will always be available, and any unavailability exceptions can be treated as serious errors rather than as part of the normal processing flow. Of course, applications can always perform actions that are not protected by promises, but resource changes that violate promises will be detected by the promise manager and undone in order to honour the guarantees it has made.

Promises are an abstract way for a client to specify the resources they need to ensure that they can complete successfully. A granted promise guarantees that the requested resources will be available when needed by later actions, but does not necessarily guarantee that any particular instance of the resource will be used to meet this promise. For example, a client may request a promise that a $5^{th}$ floor room will be available on the requested date. The response to this promise will be that a room matching the requirements will be available, not that the client has been assigned room 512. The messages and services used in the application have to reflect this level of abstraction, in this case by later making a booking for a $5^{th}$ floor room, rather than trying to confirm a booking for room 512.

Promises are both a pattern and a protocol that supports this pattern. The pattern is simply that client applications determine the constraints they need to have hold over a set of resources and express these as predicates that are sent within promise requests to a promise manager. The promise manager will consult with resource managers to determine whether a promise can be granted, and reply with either a granted or rejected response. Once a promise has been granted, the client application can continue and call services that will make changes to the resources protected by its promises with the guarantee that they will be successful if they are within the constraints implied by its promises. Client applications then release their promises by sending promise release messages to their promise managers. Promise release requests can be combined with application request messages. In this case the promise release and the application request form an atomic unit, and the promise will only be released if the associated action succeeded.

The Promises model places no limitations on the nature or form of predicates, nor on the way that promise managers should implement these predicates to guarantee that they hold despite concurrent updates to the same resources. This flexibility means that promise managers and resource managers are free to implement what ever form of constraint checking or isolation mechanism is best for the type of resource being protected.

Some forms of promises could be implemented using the common business practice sometimes called 'soft locks'. This approach uses a field in the database record to show whether an item has been allocated or reserved for a client. The record is not locked against access once the allocation has been made; instead applications read this field when looking for available resources and ignore any record that has been already allocated. Different forms of promises, such as guaranteeing that there will be enough money in an account to pay for a future purchase, could best be implemented using techniques such as escrow locking [8].

The Promise pattern accommodates both of these ways of implementing isolation, but it is more general, separating the model and its supporting protocol from any specific implementation or resource schema considerations. The flexibility that results lets us also support more general predicates where the actual allocation of a particular resource to a client is delayed to long after the promise is made, and also to support promises over pools of different but acceptable resources that export the same set of properties. Section 5 discusses a range of implementation alternatives.

The motivation behind the development of the Promises approach to isolation was to provide application programmers with something akin to the simplicity that comes from the traditional ACID transaction model. By implementing weaker but effective constraints over shared resources, we wanted to let programmers establish those resource-based pre-conditions needed to ensure their application can complete successfully, letting them then write their application code with the guarantee that concurrent activities could not violate these promises. Promise violation is still possible for other reasons (an accident might damage previously-promised stock or a third party may default on a promise they have made) but these incidents can now be treated as serious exceptions. This is very far from the situation without isolation where the effects of concurrency are common enough that they need to be included throughout the normal processing paths.

The promises obtained by clients conceptually place constraints on the behaviour of the services that they invoke. Clients get promises about resource availability and the services they then call should only make changes to protected resources that comply with these promises. For example, if a client obtains a promise that 5 pink widgets will be available to fulfil an order, then the services it calls can complete the order process for these promised goods, or the client can release the promise. The client should not use the promise for pink widgets to ask the order service to deliver some un-promised blue widgets. This restriction on the behaviour of services could be largely theoretical, being more like a design pattern than a type-safety mechanism, or the restrictions could be enforced to some degree by promise and resource managers.

Our proposed Promise protocol fits very naturally into the SOAP protocol and the Web Services model. All of our promise protocol messages can be transferred as elements in SOAP message headers and the associated actions can be carried within the body of the same SOAP messages. The fit between the Promise protocol and SOAP is discussed more fully in Section 6.

We are not the first to propose transaction-like models based on conditions that must be preserved and Section 9 points to previous work in this area. Our key innovations lie in the analysis of the variety of resources and conditions, in considering how to atomically combine several related aspects of managing a single promise, and in integrating these ideas into the services-oriented message exchange framework.

## 3. RESOURCES AND PREDICATES

This section discusses several different ways that resources can be viewed by client applications, and how these differences are reflected in the types of predicates that can be used in promises over the availability of these resources. Applications can use these different types of resource availability predicates to obtain just the degree of isolation they need for their purposes, without needing to resort to using traditional locking techniques.

Predicates are simply Boolean expressions over resources. Our model imposes no restrictions on the form these expressions can take, and in practice their form will depend on the application involved, nature of the resources and the way we want to view these resources at the time.

The simplest form of predicate expression is an application-dependent request for resources, such as asking for 'room 212, Sydney Hilton, 12/3/2007'. In this case there is a close coupling between the application, the promise manager and the resource schema, and the promise manager is responsible from translating from this application-dependent predicate to any necessary queries and updates on the room availability data held by the resource manager. The relationship between predicates, applications and resources can be much more abstract than shown in this simple example, and complex applications could define their own resource predicate language and implement their own promise managers to guarantee resource availability.

In their most general and complex form, predicates can be general Boolean expressions over defined resource availability data that is specified using standard schemas. In this case, the client would be responsible for understanding resource schemas and how resource availability is represented, and for constructing suitable predicates in the agreed standard syntax. The promise manager in this case can be completely general purpose, knowing nothing about the applications, schemas or resource availability. All that the promise manager has to be able to do is maintain sets of predicate expressions represented in this standard syntax, check them for consistency, and evaluate them with the assistance of the appropriate resource manager. For example, we could send and maintain resource availability predicates written in a standard language such as XPath or SQL, and have these query expressions evaluated by a compatible resource manager whenever the promise manager needs to check for resource availability or predicate violation.

Predicates are expressions over resources but the form and structure they take in any particular application can depend on the way we regard the resources involved. Different applications may want to treat the same physical resource, such as a particular airline seat or an individual pink widget, in different ways, and so will want to use different types of predicates to achieve the required level of isolation from any other applications that might be using the same or related resources at the same time.

In this section we discuss three different ways of regarding resources: *anonymous view, named view,* and *view via properties*. These abstractions were derived from a study of different isolation mechanisms commonly used in existing business practices. These different ways of viewing resources influence the sort of predicates that clients will need to use in order to achieve the level of isolation they require to always operate correctly.

### 3.1 Anonymous View

From the point of view of client applications, some resources can naturally be regarded as pools of indistinguishable and identical resource instances, any of which could meet a client application's requirements. All the resources in the same pool have the exactly same values for the set of attributes that are relevant to the client and it is not important to the client which items from the pool it is allocated and when this allocation takes place.

Most retail goods can be regarded as anonymous for many purposes. Barnes and Noble may have many copies of each book title in stock, and a client who wants a promise that a book will be available does not care which physical copy they are given when the order is dispatched. In this case, the book title represents a resource pool, consisting of many identical and indistinguishable copies, and all that the retailer needs to track in order to be able to make promises about availability is the number of copies they have available for sale.

Financial applications, such as banking, use anonymous resources all the time. For example, if a promise is made that a client application will be able to withdraw $500 from an account, the bank is not obliged to set aside five specific $100 bills, uniquely identified by their serial numbers.

There can be any number of promises outstanding on anonymous resources, the only constraint being that the sum of all promised resources should not exceed the resources that are actually available. For example, our bank can grant many promises against Alice's account, just as long as the account will not be overdrawn if all of these promises are followed by withdrawal requests.

The availability of anonymous resources is usually explicitly tracked and recorded in an attribute associated with each resource pool. These attributes are traditionally called something like 'quantity on hand' or 'account balance'.

### 3.2 Named View

Clients using a named view of a resource know that each instance of the resource is unique and possesses an identifier, such as a serial number or some other set of distinguishing characteristics that can be used to refer to it,. Clients can obtain a promise about the availability of a resource based on this identifier, and they can later make use of that resource instance, knowing that the promise will ensure it will be available when needed.

Some resources are naturally unique and there is only one instance of a given resource. For example, used cars could be considered unique and not interchangeable, as each one is distinguishable by the distance it has travelled and its condition. A client who gets a promise on a particular vehicle is expecting to get that one, not an 'equivalent' substitute. Conversely, new cars and hire cars would normally be accessed anonymously by model or category as they can be considered identical for the purposes of selling or hiring.

Resources such as airline seats or hotel rooms are another common class of named resources. These are virtual resources which represent the opportunity to use a (more or less) physical resource at a specific time. For example, 'Room 212, Sydney Hilton', 12/3/2007' names a specific room instance, and the date is the necessary part of the unique identifier that distinguishes one booking for the room from another.

The concepts of named and anonymous resources are about the way client applications view the resources, not about the resources themselves. A group of related named resources might be accessed anonymously in some situations, and by their unique names in others. For example, each seat on a flight has a unique name (e.g. seat 24G on QF1 departing on 8/10/2007). Some client applications may let customers try to book specific seats on a flight, and so need named access to the seat instance. In many cases though, all economy seats will be regarded as equivalent, and client applications will be using anonymous access to get promises about the availability of economy class seats on that flight.

The availability of named resources will often be tracked by the use of something like free/busy attributes associated with each resource instance. Many resources will support both anonymous and named views at the same time, allowing some clients to obtain promises on specific resources instances while others are getting promises over a collection of such resource instances.

A single named resource instance cannot be promised to more than one client application at the same time, regardless of the predicates being used and how resources are being viewed by client applications. For example, if one client is promised 'seat 24G on QF1 departing on 8/10/2007', this seat must not be included in the considerations leading to the granting of a promise for an arbitrary economy-class seat on the same flight.

### 3.3 View via Properties

The concepts of named and anonymous resource views we just discussed are really based the properties (or attributes) exposed by a resource, and the characteristics of these properties are what determine the type of promise predicates can be requested over these resources. If a set of properties can be used to always uniquely determine a specific resource instance, we can use these properties in predicates where we want a named view of the resources. If a set of properties inherently determine a set of resource instances, then we could use these properties when we want anonymous access to a pool of acceptable and interchangeable resources.

An individual resource or collection of resources would normally expose multiple properties, many of which could be of interest to clients and potentially be the target of promise predicates. For example, a hotel booking service would maintain a collection of rooms and information about their availability on specific dates. Each of these rooms has a number of properties, such as the size and type of beds, whether or not smoking is allowed in the room, whether or not there is a view, and which floor it is on. All of these properties can be used in promise predicates by client applications wanting to determine room availability.

Different client applications, acting on behalf of different customers, can make concurrent requests over the same collection of rooms and use different sets of these properties in their promise predicates. For example, one customer may be asking for a room with a view, while another might be requesting any $5^{th}$ floor room.

Room 512 could be a suitable available resource that would allow the promise manager to grant either of these requests, but the manager has to ensure that the same room is not allocated to both requests at once. The use of different properties in the two competing promise requests makes this task more difficult as it may not be straightforward to see that their predicates are effectively overlapping.

Users may regard some properties as essential and others as desirable but not required, and this could be reflected in their promise predicates. The interplay between essential and desirable properties when obtaining a promise may be complicated and could lead to systems where the promise requestor and the promise maker negotiate to find a promise that is both satisfiable and maximally desirable. For example, the client may initially request a non-smoking room with a view and twin beds, and eventually accept a promise for a room with just twin beds.

Another interesting possibility is that the values of certain properties could be treated as ordered in acceptability, with it being understood that a promise can be satisfied either by a resource that meets the precise value for a property as requested or by one offering a 'better' value. For example, a customer who holds a promise for an economy class airline seat will not normally complain if, when they fly, they are upgraded to business class.

Predicates are expressions over the values of abstract properties of resources, not over concrete fields in database tables. This abstraction gives rise to the possibility of treating resources polymorphically, allowing a single predicate to cover any number of acceptable resources as long as they all expose the required properties. For example, a hotel booking service could aggregate availability information from a number of providers, each with their own schemas for describing available rooms. A single predicate could be used to obtain a promise from any of these providers, as long as they all exported the set of properties required by the predicate (or if the properties they do export can be transformed to the required ones by the promise manager).

## 4. ATOMICITY AND PROMISES

In this section we identify three important atomicity requirements for the implementation of promises and promise managers. While the autonomy of service-providers means that there is no way to demand atomicity across long duration business processes, it is feasible to require that specific atomicity guarantees apply during the handling of a single Promise message. These requirements are:

*Request guarantees on several predicates at once.* While it may be common to seek a single guarantee such as 'ensure that at least 5 widgets are available when I decide to buy them', sometimes a client will want to ensure that several different properties (perhaps involving several resources) will all be true when the resources are required at later stages of the application's execution. The classic example is from travel planning, where a client may want a promise that a flight and a rental car and a hotel room will all be available. By treating the evaluation and granting of all the predicates carried in a single promise request as an atomic unit, the client can ensure that they will either get all the resources they need or none of them. As an aside here, the travel agent client could also build up the set of required promises needed by obtaining them one at a time, trying alternative resources and predicates when other promise requests are rejected.

*Perform an action which depends on, but violates, a previously promised condition, together with releasing the promise.* One common pattern where promises are useful is where a promise of resource availability is used to protect a later operation which consumes the resource (and thus makes it not available any more). Suppose an art gallery service has promised a client that a particular painting will be available, and the client then goes ahead and buys the painting. When the purchase occurs, the gallery service is released from the promise (the client cannot expect the painting to still be available after they themselves bought it!); however if the purchase fails for some reason (perhaps no shipper is available that day) then the promise should remain in force. In this case, the promise release and the action which depends on the promise form a unit and both parts must succeed or fail together.

*Modify the predicate whose preservation is promised, by obtaining a new promise and releasing a previous one atomically.* An important use-case is where the client requests changes to promises they have already been granted. The requested change can be to upgrade the promises, or to weaken them. For example, if a client has obtained a promise that an account will have a balance of at least $100, they may find that their anticipated later withdrawal has changed to $200 (a stronger promise is needed) or to $50 (a weaker promise). In either case, it would be too restrictive to force the service to honour the new guarantee as well as the previous one, nor would the client want to release the previous one until the new one was obtained. Thus obtaining a new promise should be atomic with releasing the old one, and the previous one should be retained if the service can't guarantee the modified request.

## 5. IMPLEMENTATION TECHNIQUES

The Promise Pattern we are proposing allows clients to ask a service to guarantee that a supplied predicate will remain true for some specified time into the future. The usefulness of this proposal depends on the existence of mechanisms which will allow the provider to guarantee that they can honour these promises, regardless of other promise requests that may be made and any other actions that may take place against the same set of resources. In this section we describe several well-known techniques that could be used in the implementation of promises. Some of these techniques have been used in a proof-of-concept implementation [6] that is discussed briefly in Section 8.

These implementation techniques are not meant to be exposed to clients through the language used to express promise predicates. This principle means that clients can express their resource requirements by using abstract predicates over resource properties, and the promise manager that receives these requests can then use whatever techniques it wants to implement the promises and meet the guarantees it has made. This approach lets the client deal in the abstractions of predicates and resources, and gives the promise manager the ability to implement these abstractions in whatever way is best at the time, and to change these implementations over time without forcing corresponding changes in client applications.

- Resource Pool: In managing anonymous interchangeable resources, it is common to keep the available instances of each resource in a pool, and move them to a separate 'allocated' pool to ensure that a promise can be honoured. For example, when we promise that we can supply 10 widgets, we remove 10 widgets from the pool of available widgets and place them in the allocated pool. The digital equivalent can be implemented by keeping a count of available and allocated items in the record corresponding to each type of resource. This technique is similar to escrow locking [8].

- Allocated Tags: In the case of resources that are accessed via a named view, we can keep an availability status field as part of the data used to describe the resource instance. This field would be set to something like 'available' initially and then to 'promised' when the instance was provisionally allocated to a client as a result of making a promise. It would then be either set to 'taken' by a subsequent action, or would be reset back to 'available' if the promise is released and the client has no further use for the resource.

- Satisfiability Check: The promise manager keeps a record of all the promises it is currently committed to honouring and also has access to the current state of all resources covered by these promises. Whenever a new promise request is received, the manager checks that it and all relevant existing promises can be honoured, based on the current state of the resources involved. Similarly, a check is performed after every client-requested operation has completed to be sure that the state afterwards still allows all existing promises to be honoured.

If property-based access is used, the decision about which resource will be used to honour a granted promise can be delayed until the execution of the operation which takes the resource. In this approach, the promise manager needs to be able to check the compatibility of a set of promises with the state of the resources. This might be done by finding a matching in a bipartite graph where edges link the untaken resources to the promise predicates that they can satisfy.

One consequence of this model is that the availability of a resource is indicated by the presence (or absence) of a covering predicate, as well as (possibly) fields in the resources themselves. In contrast to the 'allocated tag' mechanism just described above, we now have the situation where the availability field in the resource now only indicates whether or not the resource has been definitely taken. This means that status information for a single set of resources is now distributed between the promise and resource managers, and special care will be needed to ensure consistency.

- Tentative allocation: This is a hybrid mechanism, where property-based promise requests are met by marking the chosen resource instances as 'promised', and also remembering the specific predicate that resulted in this resource allocation. If a later promise request is not satisfiable from the pool of unallocated instances, the manager can consider rearranging these tentative allocations to allow it continue to meet all previous promises as well as granting the new request. For example, a request for a hotel room with a view may lead to tentatively allocating room 512 (on the basis that it has a view). When a later request is made to promise a $5^{th}$ floor room, the system may reallocate 512 to the new request as long as a different room with a view can be still be provided to meet the earlier request.

- Delegation: Promises are made that rely on the promises of third parties. For example, a purchase order can be accepted by the merchant if it has received a promise from the distributor that a backorder will be fulfilled on time. In this scenario, the promise is delegated from the merchant to the merchant's supplier.

As mentioned earlier, the architectural model we are using here has promises being granted and guaranteed by a Promise Manager. This system component acts as an intermediary between clients and services by receiving and granting promises, working with resource managers to help determine availability and passing application requests on to services for execution.

In this model, client applications always send both promise messages and application requests to an intermediate promise manager rather than directly to services or resource managers. The promise manager will act on the promise messages, consulting with applications and resource managers as needed to determine if promises can be granted. Application requests pass through the promise manager so that they can be rejected if any associated promises cannot be granted or if executing the request would cause existing promises to be violated.

This is only a conceptual model, although it is the one implemented in our prototype. Actual implementations are free to implement the required promise functionality in any way at all. Implementations could move all promise functionality into the application services, letting them use whatever application-dependent mechanisms they wish to express predicates, record promises and determine resource availability. Another alternative would be to move the responsibility for granting and enforcing promises to the resource managers where they could be implemented as a form of dynamic integrity constraint.

## 6. PROMISE PROTOCOL

This section discusses the structure of some protocol elements that could be used in a SOAP-based implementation of the Promise Pattern. In this protocol, clients and promise managers exchange promise-related information using <promise> and <environment> message header elements. <Promise> elements are used in the creation and release of promises. <Environment> elements are used to specify the promise context that applies for the SOAP service requests carried in the associated message body.

A <promise> element can have zero or more <promise-request> elements; each representing one request for the recipient to make a promise that will guarantee the included predicates for a certain period of time. A <promise> element can also include zero or more <promise-response> elements which are used to return outcomes from previous requests that flowed in the reverse direction. Each participating service can act as both client and promise-maker, so a single <promise> element can include both <promise-request> and <promise-response> elements.

A <promise-request> defines:

- A *request identifier* that can uniquely identify each promise-request. This request identifier is used to correlate promise-requests and promise-responses.

- A set of *predicates* that specify the conditions on which the client will rely in a later interaction and that the promise-maker must maintain.
- A set of *resources* that specify the subjects of the promise.
- A *promise duration* that indicates how long the client wants the promise to be kept.
- An optional set of *promise identifiers* that refer to existing promises that can be released if this new promise request is successfully granted.

Each promise-request must be treated atomically. All of the predicates over the specified resources must be promised or the entire promise must be rejected. A promise request may hand back previous promises in exchange for new promises, and if these new promises cannot be granted, the existing promises must continue to hold.

Promise makers send promise responses back to promise requestors to inform them whether their promise requests have been accepted or rejected. The elements of a <promise response> are:

- A *promise identifier* that the promise maker uses to uniquely identify this promise.
- A *promise result* that says whether a promise request is accepted or rejected. Promise responses could also return other results, such as 'pending' or 'accepted with the condition XX' but these possibilities have still to be investigated.
- A *promise duration* that indicates how long the promise manager will guarantee to keep this promise. This may be the same as the duration which was requested, but the promise manager might, for example, offer a guarantee that expires sooner than the client wished.
- A *promise correlation* which is the *request identifier* of the earlier promise request.

Successful promise requests establish promise environments. Application requests can specify that they must be executed within a specific promise environment (with the set of resource guarantees defined by its promises) by including an <environment> element in the associated message header. An <environment> must define;

- A set of *promise identifiers* that define which promises will apply for the execution of the request.
- A corresponding set of *promise release options* that indicate whether the associated promises should be released after the request has completed.

We note that each message may contain any subset of the different elements relating to promises, and these may be related to the message body or unrelated. For example, we allow an application message from A to B to contain a related request for B to make a promise, and it can also carry a piggybacked response reporting on the outcome of a previous request that B had sent to A.

## 7. PROMISES AND ISOLATION

The key contribution of the Promise pattern is that it allows a client to check for the availability of resources and then later make service requests with the assurance that these operations will not fail because the required resources are no longer available (except for very rare catastrophic situations that might need human intervention). Programmers are relieved of the need to consider the frequent but unwelcome situation where concurrent activity has changed the truth of relied-on conditions after they were checked.

We will illustrate how applications can use promises to achieve the precise degree of isolation they require through two examples based on the merchant example mentioned earlier. Both of these examples make use of the Promise Pattern but differ in the resources involved, the way they view them and the predicates they use.

The first example [Figure 1] shows how the ordering process can check for the availability of goods using a promise and then be guaranteed that these goods will continue to be available for purchase, regardless of any concurrent activities, until the order is completed or abandoned. In this example, the customer is trying to order 5 pink widgets. As our customer doesn't care exactly which 5 of the many identical pink widgets in stock they will receive as a result of this order, we will use the anonymous access view defined in Section 3.1 for this example.

**Order process**                                                    *Promise manager*

Determine we need 5 pink widgets to be in stock
Send promise request that (quantity of 'pink widgets' >= 5)
    *Check stock levels of pink widgets and…*
    *Accept promise if >=5 currently available*
        *Record promise as predicate over stock*
        *levels, guaranteeing that at least 5 units*
        *will always be available. This predicate*
        *will be checked before any further*
        *promises are granted or purchases are*
        *performed.*
    *Send 'accept' <promise response>*
    *Reject promise request if <5 units available*
        *Send 'reject' <promise response>*
If promise rejected
    Terminate order process saying goods unavailable
If promise accepted…
    Continue processing order (organise payment, shippers)

Send 'purchase stock' request to promise manager
and release promise to keep stock level >= 5
    *Pass 'purchase stock' to application service*
    *(Release 5 pink widgets for delivery*
    *Reduce stock-on-hand by 5)*
    *Remove this promise from the set of*
    *predicates over the pink widget stock level*

**Figure 1. Outline of Ordering Process Code**

The second example is more complex and illustrates the flexibility of promise predicates. In this example, our merchant offers 'next day' shipping to its customers for a fixed additional cost on all orders. The order process asks the promise manager for the shipping component for a promise of next day delivery, with the predicate making no assumptions about how this promise will be implemented or needing any information about the structure of the shipping component and its internal states. The shipping promise manager could implement the promise by obtaining soft-locks on warehouse and shipping capacity but other implementations are possible. The merchant may even have a number of shipping alternatives available, each with different capacity and cost structure, and the actual choice of which shipper to use could be deferred until shipping is required in order to reduce costs and optimise utilisation. This flexibility is not visible to the order process or the customer, all that they need to know is that the shipping component has promised next-day delivery and guarantees that this will occur.

## 8. PROTOTYPE IMPLEMENTATION

We implemented a prototype Promise-based system as a proof-of-concept demonstrator and to help further explore some of the design concepts and issues involved. This prototype is more fully discussed in [6]. The overall architecture of this system is shown in Figure 2. The implementation follows the conceptual model discussed earlier, with the promise manager being a separate component, and uses a satisfiability-based mechanism for checking promises. The messages sent by the client to the promise manager can include both Promise and Action parts, keeping with the protocol model discussed in Section 6.

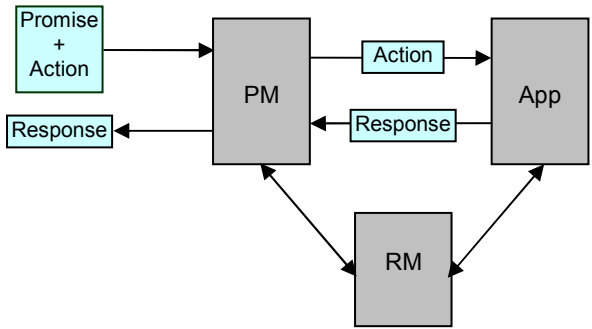

**Figure 2. Structure of Promise Prototype**

The prototype Promise Manager is best seen as an intermediary between the client and the application. The client adds promises header messages to its normal service requests and sends them to the promise manager for processing. The promise manager then does its work and passes the request on to the application. The roles of the components in this promise system are:

- *Promise Manager (PM):* The promise manager receives each message as it arrives from the client and breaks it up into its Promise and Action component pieces. If a message contains a Promise part, this is split into its promise request and promise environment parts and any new promise requests are checked for consistency against the existing promises and resource availability. After this step, any Action is passed on to the associated application and the promise manager waits for a response. If the Action succeeds, the promise manager then uses the supplied promise environment to update the set of applicable promises and checks once again that all relevant promises are still consistent with the resource availability information held by the resource manager. This step is what allows the promise manager to guarantee that promises will be honoured, regardless of what state changes have occurred as a result of executing the Action. If all promises can still be honoured, the promise manager passes back the response it received from the application back to the client. If the result of the action was that promises were violated, the promise manager will roll back the changes made by the Action and return a failure message to the client. In the prototype, an ACID transaction is used for the complete processing of each request, and this allows us to either commit or rollback any changes made by the application after checking for promise violations.

- *Application:* The responsibility of the application is to process the action request passed from the promise manager. The application uses a resource manager to keep the global system state which is shared between operations. After the action has completed, the application sends a response message back to the promise manager.

- *Resource Manager (RM):* The role of the RM is to store the state of the system, and to process queries and updates on this data as requested by the application and the promise manager.

The most critical part of the promise manager is the code that guarantees the validity of non-expired promises by ensuring that sufficient resources are available to satisfy every active predicate.

The promise manager keeps a record of all non-expired promises and their predicates in a 'promise table'. Promises are placed in this table when they are granted and removed when they are released. The promise manager evaluates incoming promise requests by checking that the new predicates do not conflict with any existing promises and that they are consistent with the current state of the resources involved. This process of evaluating a set of promises for consistency is called 'promise checking'. The actual code used for this checking depends on the type of resource view embodied in the predicates used in the promises.

For the case of a named resource, promise checking is relatively simple and we just have to ensure that one of the following situations holds: there are no duplicate promises for the resource (as identified by its unique identifier); or the resource must be recorded as available in the RM, and there is at most one unexpired promise over that resource.

For an anonymous resource where there is a pool of equivalent items, the promise checking process sums the quantities of the specified resource required by all unexpired promises, and this value must be at least as large as the amount recorded in the RM as being available.

Property-based views of resources are much more complicated because deciding whether to grant promise requests requires bipartite graph matching. Checking promises over these views is not implemented in our prototype at present.

Promise checking is used in several places in the promise manager

- *Making New Promises*: Granting a new promise must consider the mutual satisfiability of all existing unexpired promises and the requested promise, using currently available resources as known by the RM. The request will be granted if this consistency check passes, and rejected otherwise.

- *Executing Actions*: The Application executes actions that were coded without explicit knowledge of the PM or its promises. These actions might change the state of resources, for example by updating the account balance upon receiving payment or modifying the availability of rooms when customers make a booking. In a well-designed system, actions would make no state changes except those that were guaranteed by relevant promises. However the promise manager cannot rely on the application code being always well-behaved, so the promise manager also has to check for consistency after an action has been completed. This ensures that the state changes made by the application have not violated any unrelated promises. Applications are allowed, of course, to make state changes that will violate those promises that are being released atomically with the action.

- *Updating Existing Promises*: Promise clients can request to update existing promises. Updating existing promises can be seen as the atomic combination of two operations: removing the previous promises and creating new promises. The promise manager has to check the consistency of the proposed new set of promises and current resource availability.

Information about promises and resource availability are stored in different places and controlled by different managers, but they are both accessed as part of promise operations and have to be consistent. For example, granting a promise request involves examining the state of resources held in the RM and examining the predicates held in the promise table, as well as inserting the new promise into the promise table. Without taking special care when coding the promise manager, we could have been vulnerable to race conditions and other isolation failures resulting from concurrent promise operations.

The solution we adopted here was to wrap each promise operation in a transaction. This transaction is started when we begin processing each client request and committed or rolled back just before the result of the request is returned to the client. This transaction covers all of the action code executed inside the application as well as the subsequent promise checking code (including modifications to the promise table). This means that all accesses to the resource manager, as well as changes to the promise table are transactional, and this gives us the required level of isolation between concurrent activities. Note that the transaction is local to a trust domain and short-duration. It does not include any external messaging or code outside the scope of the service and its associated promise manager.

## 9. RELATED WORK

One of our inspirations in this project was the early ConTract work of Wachter and Reuter [11]. This introduced the importance of expressing preconditions ('entry invariants') needed to allow actions within a workflow to execute successfully. The authors identified several different styles of ensuring that these preconditions still hold at the time when applications rely on them later in an execution. Among the styles proposed was the use of semantic locks to preserve conditions and notifying the client when a checked condition changes. Our work extends the semantic lock ideas of ConTract to the services world with its interacting autonomous participants. Our consideration of atomically combining steps is also new. We provide a richer analysis of the variety of resource and predicate types, and of the ways to ensure that predicates remain true over an extended period. We also support a variety of possible implementation mechanisms, each tailored to the needs of specific ways of viewing and accessing resources.

In previous work [7], one of us developed a transaction model for spatial data which was based on explicit constraints that could be set and unset to limit concurrent modification of properties of the data. Our current paper extends this to a world of autonomous services; as well we now offer an analysis of predicate types, and a better mechanism to structure the operations by providing atomicity between aspects of a single step of the promise exchange.

Recently Dieter Gawlick and other members of the Grid Computing community have suggested the 'Option' protocol [2] for reserving access to resources. This has similarities to Promises but our work deals with a wider class of conditions including those on anonymous resources and property-based views of resources, and supports a wider choice of implementation mechanisms. Also, our use of atomicity allows us to unify concepts such as securing, modifying, confirming, and dropping which are represented as separate message types in [2]. The "options" approach has been implemented inside an Oracle database management system, using "data cartridges" to define data types with appropriate indexing and triggers (D. Kossmann, private communication).

The idea of an organisation making a promise about future performance or behaviour is quite common in bricks-and-mortar businesses, and most of the implementation mechanisms we considered have long precedents in business practice. For digital data, many implementation techniques have been proposed which offer the effect of promise keeping. Conventional database locking provides the semantic effect of ensuring that data is not altered between the time a condition is checked and the time it is needed, despite any concurrent activities, but the locking mechanism assumes an environment where activities run very quickly and all participants can be trusted to hold locks. These assumptions are inflexible and not suited for data under high contention or for today's service-based applications. Alternative mechanisms have been developed within database engines for allowing higher concurrency based on knowledge of the semantics of the data. For example, escrow locking [8] deals with numeric data under operations that add or subtract, by recording high and low limits for the possible values, while granular locks and predicate locking have been proposed as a means of preventing phantoms [1]. The implementation techniques available for promises are similar to these, but there are significant differences. Promises have a limited duration, so a promise maker is not surrendering site autonomy to an extent that would be unacceptable given the limited trust assumptions typical of cooperating parties. Also, because unfulfillable promise requests are rejected immediately rather than blocking, we do not have to worry about the deadlock issues that plague lock-based algorithms.

There are interesting parallels between promises and the IMS/VS Fast Path mechanism [3]. In Fast Path, each operation is structured as a predicate check and a transformation on the data. The predicate is checked when the operation is submitted, and then at commit-time, the check is repeated, and the transformation is performed (provided the check succeeded). We can consider the operation submission as like a promise request, and commit as like the operation done under promise protection; however, in Fast Path, other operations do not worry about outstanding predicates, and so the commit check might fail because of concurrent activity.

Promises are also analogous to integrity constraints, and many researchers have considered how to enforce integrity in database management systems. In seminal work, [10] showed how one could enforce integrity by modifying update statements, and [9] showed how compile-time checks could ensure that application code preserved constraints. Techniques like these might be useful in implementing a promise manager which needs to check each client action for compatibility with previously granted promises. However, there are important differences between integrity constraints and promises. Most significantly, each integrity constraint can be considered independently, while promises need to be satisfiable by disjoint resources. For example, two integrity constraints 'balance>100' and 'balance>50' are both met if the balance is 120, but two promises for 'balance>100' and 'balance>50' imply that the balance must be kept over 150. With property views, promise satisfiability can require a graph matching algorithm, whereas integrity satisfiability is just logical satisfiability.

Our Promises pattern unifies and abstracts over many possible implementation mechanisms, including those that are based on previous work mentioned above. The Promises approach offers a common way for clients to work without knowledge of the implementation technique used inside a service that can maintain some property between the time it is checked and a later time when the client relies on the property.

## 10. CONCLUSION

In this paper we propose a unified approach to describing the interactions between a client and a service where the client can make sure that some condition over resources will hold at a later time, despite concurrent activities that occur between the check and the use of the condition. We have analysed the variety of resource types and conditions on those types, identifying an important distinction between resources which are accessed anonymously (where the key property is just whether a given amount or volume is available), resources which are accessed by name, and a wider class where access is based on values for some subset of a collection of properties. We have identified important cases where several promise-related activities need to be combined into an atomic unit in order to support valuable use-cases such as upgrading or weakening a previously obtained promise.

In future work, we will implement support for Promise interactions in several service-provision frameworks, including our own GAT engine [5] and also some commercial approaches. This will involve developing further implementations for checking predicates against resources, as discussed in Section 5; as well as providing simple heuristics to choose an appropriate implementation technique for each class of resources. We also will integrate the processing of promises with other frameworks for service-oriented messaging, including the transaction support found in standards like WS-BusinessActivity.